\documentstyle[11pt,newpasp,twoside,epsf]{article}
\reversemarginpar

\newcommand{\be}{\begin{equation}}
\newcommand{\ee}{\end{equation}}

\begin{document}
\title{Singular Isothermal Disks \\ and the Formation of Multiple Stars}
\author{Daniele Galli}
\affil{Osservatorio Astrofisico di Arcetri, Firenze, Italy}
\author{Frank H. Shu}
\affil{Department of Astronomy, University of California, Berkeley, USA}
\author{Gregory Laughlin}
\affil{NASA/Ames Research Center, Moffett Field, USA}
\author{Susana Lizano}
\affil{Instituto de Astronom\'{\i}a, UNAM, M\'exico}

\begin{abstract}
A crucial missing ingredient in previous theoretical studies of
fragmentation is the inclusion of dynamically important levels of
magnetic fields.  As a minimal model for a candidate presursor to the
formation of binary and multiple stars, we therefore consider the
equilibrium configuration of isopedically magnetized, scale-free,
singular isothermal disks, without the assumption of axial symmetry.
We find that lopsided ($M=1$) configurations exist at any dimensionless
rotation rate, including zero.  Multiple-lobed ($M$ = 2, 3, 4, ...)
configurations bifurcate from an underlying axisymmetric sequence at
progressively higher dimensionless rates of rotation, but such
nonaxisymmetric sequences always terminate in shockwaves before they
have a chance to fission into separate bodies.  We
advance the hypothesis that binary and multiple star-formation from
smooth (i.e., not highly turbulent) starting states that are
supercritical but in unstable mechanical balance requires the rapid
(i.e., dynamical) loss of magnetic flux at some stage of the ensuing
gravitational collapse.  
\end{abstract}

\section{Magnetic Fields in Star Forming Clouds}

On scales larger than small dense cores ($\sim 0.1$~pc), magnetic
fields are more important than thermal pressure (but perhaps not
turbulence) in the support of molecular clouds against their
self-gravitation (see the review of Shu, Adams, \& Lizano 1987).
Mestel has long emphasized that the presence of dynamically significant
levels of magnetic fields changes the fragmentation problem completely
(Mestel \& Spitzer~1956; Mestel~1965a,b; Mestel~1985).  Associated with
the flux $\Phi$ frozen into a cloud (or any piece of a cloud) is a
magnetic critical mass:
\be 
M_{\rm cr}(\Phi) = {\Phi \over 2\pi G^{1/2}}.  
\label{MPhi} 
\ee 
Subcritical clouds with masses $M$ less than $M_{\rm cr}$ have magnetic
(tension) forces that are generally larger than and in opposition to
self-gravitation (e.g., Shu \& Li 1997) and cannot be induced to
collapse by any increase of the external pressure.  Supercritical
clouds with $M > M_{\rm cr}$ do have the analog of the Jeans mass -- or,
more properly, the Bonnor-Ebert mass -- definable for them, but unless
they are highly supercritical, $M\gg M_{\rm cr}$, they do not easily
fragment upon gravitational contraction.  The reason is that if $M\sim
M_{\rm cr}$ for the cloud as a whole, then any piece of it is likely to be
subcritical since the attached mass of the piece scales as its volume,
whereas the attached flux scales as its cross-sectional area.  Indeed,
the piece remains subcritical for any amount of contraction of the
system, as long as the assumption of field freezing applies.  An
exception holds if the cloud is highly flattened, in which case the
enclosed mass and enclosed flux of smaller pieces both scale as the
cross-sectional area. This observation led Mestel (1965a,b; 1985) to
speculate that isothermal supercritical clouds, upon contraction into
highly flattened objects, could and would gravitationally fragment.

Zeeman observations of numerous regions (see the summary by Crutcher
1999) indicate that molecular clouds are, at best, only marginally
supercritical.  The result may be easily justified after the fact as a
selection bias (Shu et al.  1999).  Highly supercritical clouds have
evidently long ago collapsed into stars; they are not found in the Galaxy
today.  Highly subcritical clouds are not self-gravitating regions;
they must be held in by external pressure (or by converging fluid
motions); thus, they do not constitute the star-forming
molecular-clouds that are candidates for the Zeeman measurements
summarized by Crutcher (1999).  The clouds (and cloud cores) of
interest for star formation today are, by this line of reasoning,
marginally supercritical almost by default.

\subsection{Pivotal States and Self-Similarity}

The stage leading up to dynamic collapse of a magnetically subcritical
cloud core to a protostar or a group of protostars is believed to be
largely quasi-static (e.g., Nakano~1979, Lizano \& Shu~1989,
Tomisaka~1991, Basu \& Mouschovias~1994).  To describe the transition
between quasi-static evolution by ambipolar diffusion and dynamical
evolution by gravitational collapse, Li \& Shu~(1996) introduced the
idea of a {\it pivotal state} to indicate scale-free, magnetostatic
configurations just before the onset of gravitational collapse
(protostar formation and envelope infall). For these states, the
density distribution approaches $\rho \propto r^{-2}$ for an isothermal
equation of state when the mass-to-flux ratio has a spatially constant
value, a condition that Shu \& Li~(1997) and Li \& Shu~(1997) termed
{\it isopedic}.  Numerical simulations of the contraction of magnetized
clouds do show that the mass-to-flux ratio remains constant over
several decades of spatial extent (see e.g. Basu \& Mouschovias~1994).

Magnetized self-gravitating equilibria tend to be somewhat flattened,
unless high levels of toroidal fields are present. When the support
against self-gravity is dominated by poloidal magnetic fields and/or
rotation, the configuration becomes a thin disk, not necessarily
axisymmetric or time independent. Also, if turbulent support is modeled
as a polytropic or logatropic scalar pressure with a relatively soft
equation of state (e.g. Lizano \& Shu~1989, Holliman \& McKee~1993),
then all magnetostatic configurations with $\lambda\leq 1$ are highly
flattened in the direction perpendicular to the magnetic field.

Li \& Shu (1996; see also Baureis, Ebert \& Schmitz~1989) have shown
that the general, axisymmetric, magnetized equilibria representing such
pivotal states assume the form of singular isothermal toroids (SITs):
$\rho(r,\theta) \propto r^{-2}R(\theta)$ in spherical polar coordinates
$(r,\theta,\varphi)$, where $R(\theta) = 0$ for $\theta =0$ and $\pi$
(i.e., the density vanishes along the magnetic poles).  We regard these
equilibria as the isothermal (rather than incompressible) analogs of
Maclaurin spheroids, but with the flattening produced by magnetic
fields rather than by rotation.  In the limit of vanishing magnetic
support, SITs become singular isothermal spheres.  In the limit
where magnetic support is infinitely more important than isothermal gas
pressure, SITs become singular isothermal disks (SIDs), with $\rho
(\varpi,z) = \Sigma (\varpi) \delta (z)$ in cylindrical coordinates
$(\varpi,\varphi,z)$, where $\delta (z)$ is the Dirac delta function,
and the surface density $\Sigma (\varpi) \propto \varpi^{-1}$.

In a fashion analogous to the singular isothermal sphere (Shu 1977),
the gravitational collapses of SITs have elegant self-similar
properties (Allen \& Shu 2000).  But it should be clear that the
formation of binary and multiple stars could never result from any
calculation that imposes a priori an assumption of axial symmetry.  In
this regard, we would do well to remember the warning of Jacobi in
1834:

{\it ``One would make a grave mistake if one supposed that the axisymmetric
spheroids of revolution are the only admissible figures of equilibrium.''}

\section{Nonaxisymmetric Equilibria and Bifurcations}

Shu et al.~(2000) and Galli et al.~(2000) started the campaign to
understand binary and multiple star-formation by considering the
equilibrium and stability of nonaxisymmetric self-gravitating,
magnetized, differentially-rotating, completely flattened SIDs, with
critical or supercritical ratios of mass-to-flux
\be
\lambda\equiv 2\pi G^{1/2} {M(\Phi)\over \Phi}\ge 1,
\label{deflambda}
\ee
(see Li \& Shu 1996, Shu \& Li~1997).  The dimensionless mass-to-flux
ratio $\lambda$ is taken to be a constant both spatially (the {\it
isopedic} assumption) and temporally (the {\it field-freezing}
assumption).  The governing equations of our problem are the usual gas
dynamical equations for a completely flattened disk (see Shu \&
Li~1997, Shu et al.~2000), except for two modifications introduced by
the presence of magnetic fields that thread vertically through the
disk, and that fan out above and below it without returning back to the
disk. First, magnetic tension reduces the (horizontal) gravitational
force by a multiplicative factor $\epsilon=1-\lambda^{-2}\leq 1$.
Second, the gas pressure is augmented by the presence
of magnetic pressure; this increases the square
of the effective sound speed by a multiplicative factor
$\Theta = (\lambda^2+3)/(\lambda^2 + 1) \ge 1$. 
In other words, {\it the equations of motions for the isopedic, isothermal
SID are identical with those of a nonmagnetized disk except for the 
tranformations}
\be
G\rightarrow \epsilon G,~~~~~a^2\rightarrow \Theta a^2,
\ee
where $G$ is the gravitational constant and $a$ the isothermal sound speed.

Under the assumption of field freezing, i.e. keeping $\lambda$ constant
in time, Galli et al.~(2000) found that prestellar molecular cloud
cores modeled as magnetized SIDs {\it need not be axisymmetric}.  The
most impressive distortions are those that make slowly rotating
circular cloud cores lopsided ($M=1$ asymmetry, see also Syer \&
Tremaine~1996).  In particular, in the absence of rotation, the system
of equations of the problem has an {\it analytical} solution, where
iso-surface-density contours are {\it ellipses} of eccentricity $e$,
\be
\Sigma(\varpi,\varphi)={K\over \varpi(1- e\cos\varphi)},
\label{S1}
\ee
with $K$ constant and $0<e<1$.  Notice that the limit $e\rightarrow 1$
produces a semi-infinite filament with mass per unit length $2\pi K$.
For values of $e$ between these two extremes, both iso-surface-density
contours and equipotentials are confocal ellipses of eccentricity $e$.
Fig.~1 shows an example of a nonrotating SIDs with $e=0.3$.

On the other hand, bifurcations into sequences with $M =2$, 3, 4, 5,
and higher symmetry require non-zero rotation rates ($>0.7$ times the
magnetosonic speeds), as shown in Fig.~2. These values are considerably
larger that is typically measured for observed molecular cloud cores
(e.g. Goodman et al.~1993).  Although seemingly more promising for
binary and multiple star-formation, the models with $M=2$, 3, 4, 5,...
symmetries all terminate in shockwaves before their separate lobes can
succeed in forming anything that resembles separate bodies.  For these
configurations to exist at all, the basic rotation rate has to be
fairly close to magnetosonic. It is then not possible for the nonaxial
symmetry to become sufficiently pronounced as to turn streamlines that
circulate around a single center to streamlines that circulate around
multiple centers (as is needed to form multiple stars), without the
distortions causing supermagnetosonically flowing gas to slam into
submagnetosonically flowing gas. The resultant shockwaves then
transport angular momentum outward and mass inward in such a fashion as
to prevent fission.

\begin{figure}[t]
\plotfiddle{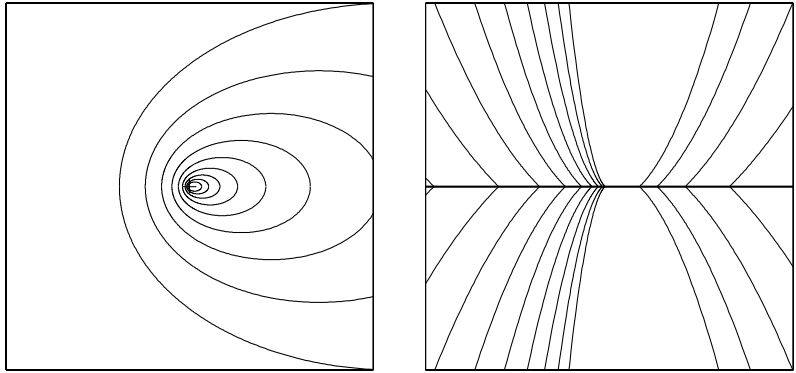}{5cm}{0}{60}{60}{-170}{-80}
\caption{({\em a}\/) Iso-surface-density contours for non-rotating SIDs,
seen face-on, are confocal ellipses with eccentricity $e$ (in this case, 
$e=0.3$). ({\em b}\/) Poloidal magnetic field lines, for the same SID
as in ({\em a}\/), seen edge-on. The magnetic field lines leave 
the disk at an angle of 45$^\circ$.}
\end{figure}

\begin{figure}[t]
\plotfiddle{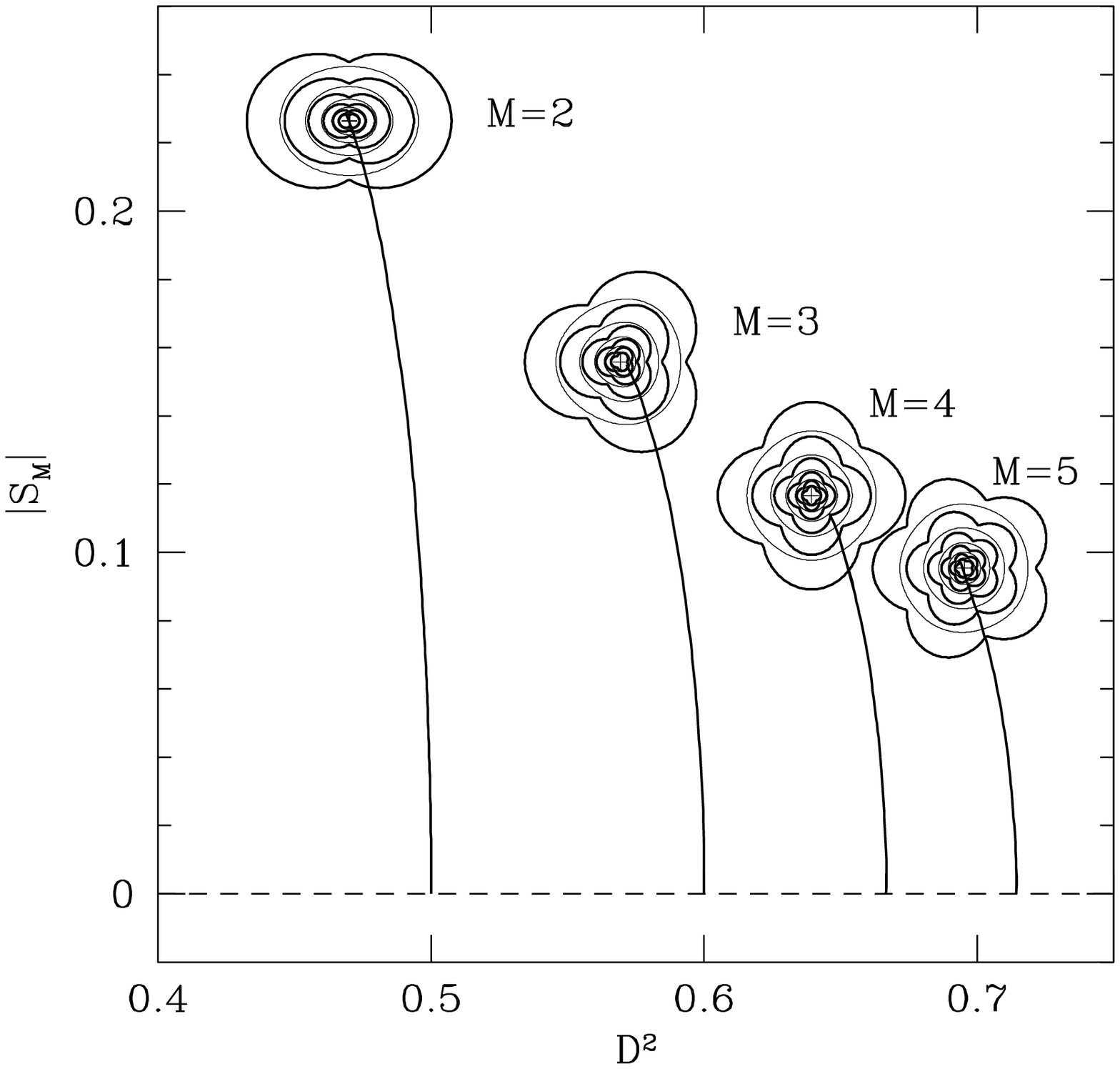}{8cm}{0}{50}{50}{-150}{-90}
\caption{Locus in the $D^2$--$|S_M|$ plane of sequences of equilibria
with given $M$-fold symmetry. Here $D$ is the ratio of the rotation
rate to the magnetosonic speed, and $S_M$ the amplitude of the dominant
component in the Fourier expansion of the surface density. The {\it
dashed line} indicates the locus of axisymmetric equilibria. Sequences
of equilibria originate from axisymmetric models and terminate because
of the occurrence of shocks. Isodensity contours ({\it thick solid
lines}) and streamlines ({\it thin solid lines}) of terminal models are
shown at the endpoints of each sequence.} 
\end{figure}

\subsection{The Need for Magnetic Flux Loss}

This negative result, combined with the analysis of the spiral
instabilities that afflict the more rapidly rotating, self-gravitating,
disks into which more slowly rotating, cloud cores collapse (also
modeled here as SIDs), is cause for pessimism that a successful
mechanism of binary and multiple star-formation can be found by either
the fission or the fragmentation process acting in the aftermath of the
gravitational collapse of marginally supercritical clouds {\it during
the stages when field freezing provides a good dynamical assumption}.

In contrast, we know that the dimensionless mass-to-flux ratio
$\lambda$ has to increase from values typically $\sim 2$ in cloud cores
to values in excess of 5000 in formed stars (Li \& Shu 1997).  Massive
loss of magnetic flux must have occurred at some stage of the
gravitational collapse of molecular cloud cores to form stars.
Moreover, this loss must take place at some point at a dynamical rate,
or even faster, since the collapse process from pivotal molecular cloud
cores is itself dynamical.  It is believed that dynamical loss of
magnetic fields from cosmic gases occurs only when the volume density
exceeds $\sim 10^{11}$ H$_2$ molecules cm$^{-3}$ (e.g., Nakano \&
Umebayashi 1986a,b; Desch \& Mouschovias 2000).  It might be thought
that cloud cores have to collapse to fairly small linear dimensions
before the volume density reaches such high values, and therefore, that
only close binaries can be explained by such a process, but not wide
binaries (McKee 2000, personal communication).  However, this
impression is gained by experience with {\it axisymmetric} collapse.
Once the restrictive assumption of perfect axial symmetry is removed,
we gain the possibility that some dimensions may shrink faster than
others (e.g., Lin, Mestel, \& Shu 1965), and densities as high as
$10^{11}$ cm$^{-3}$ might be reached while only one or two dimensions
are relatively small, and while the third is still large enough to
accomodate the (generally eccentric) orbits of wide binaries.

The linearized stability analysis and the nonlinear simulations of Shu et
al.~(2000) suggests that the collapse of gravitationally unstable
axisymmetric SIDs lead to configurations that are stable to further
collapse but dynamically unstable to an infinity of nonaxisymmetric
spiral modes that again transport angular momentum outward and mass
inward in such a fashion as to prevent disk fragmentation.  We suspect
the same fate awaits the collapse of pivotal SIDs that are
non-axisymmetric to begin with, as long as we continue with the
assumption of field freezing.  Thus, we speculate that {\it rapid
(i.e., dynamical rather than quasi-static) flux loss} during some stage
of the star formation process is an essential ingredient to the process
of gravitational fragmentation to form binary and multiple stars from
present-day molecular clouds.

\section{A Specific Example: the Molecular Cloud Core L1544}

\begin{figure}[t]
\plotfiddle{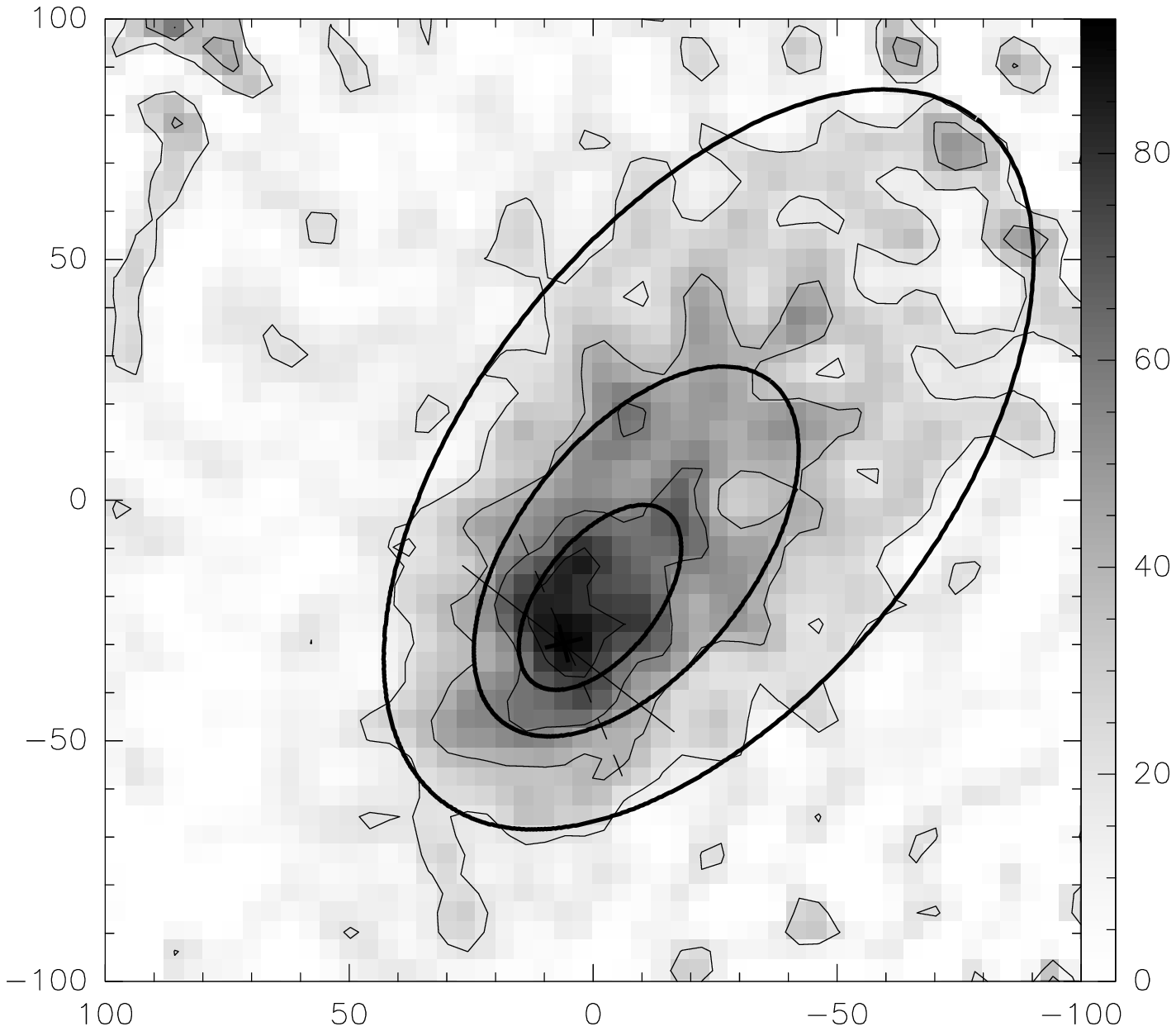}{8cm}{0}{60}{60}{-160}{-200}
\caption{Iso-surface-brightness contours ({\it thick solid lines}\/)
from a theoretically computed,
lopsided, magnetized, self-gravitating figure of equilibrium compared
with isophotal measurements of Ward-Thompson et al.~(1999) of the
submillimeter emission from heated dust grains in L1544.  The short
{\it solid line} and {\it dashed line} show the directions of predicted and
measured field inferred from submillimeter-wave polarization
observations (Ward-Thompson et al.~2000).}
\end{figure}

As an example, Fig.~3 shows an overlay of one of our eccentrically
displaced static models projected onto a map of thermal dust emission
at 1.3~mm obtained by Ward-Thompson, Motte, \& Andr\'e (1999) for the
prestellar molecular cloud core L1544.  Apart from relatively minor
fluctuations due to the cloud turbulence, the solid curves depicting
the iso-surface-density contours of the theoretical model match well
both the observed shapes and grey-scale of the dust isophotes.

Zeeman measurements of the magnetic-field component parallel to our
line of sight toward L1544 have been made by Crutcher \& Troland
(2000), who obtain $B_\parallel = 11\pm 2$ $\mu$G.  For a highly
flattened disk, which is reflection symmetric about the plane $z=0$,
integration along the line of sight yields cancelling contributions of
$B_\varpi$ and $B_\varphi$ to $B_\parallel$. The $z$-component of the
magnetic field of our model core is given by
\be
B_z={2\pi G^{1/2}\over \lambda} \Sigma.
\ee
We may now calculate the average value of $\Sigma$ within a radius $R$ as
\be
\langle \Sigma \rangle= {1\over \pi R^2}
\oint d\varphi \int_0^R \Sigma \varpi\, d\varpi=
{\lambda^2(\lambda^2+3)\over (\lambda^4-1)}{a^2\over \pi G R},
\ee
Therefore, the average 
value of $B_z$ within a radius $R$ is
\be
\langle B_z \rangle = {2\pi G^{1/2}\over \lambda}\langle \Sigma\rangle
= {\lambda (\lambda^2+3)\over (\lambda^4-1)} {2 a^2\over G^{1/2} R}.
\ee

Since we model L1544 as a thin disk with elliptical iso-surface-density
contours, its orientation in space is defined by three angles, two
specifying the orientation of the disk plane, the third giving the
position of the elliptical contours in this plane. We fix the first
angle by assuming for simplicity that the major axis of the elliptical
contours lies in the plane of the sky.  The second angle $i$ is the
inclination of the minor axis with respect to the plane of the sky
($i=0$ for a face-on disk) and can be adjusted to fit the
observations.  The third angle, specifying the ellipse's orientation in
the disk plane, is given as 38$^\circ$ north through east by
Ward-Thompson et al.~(2000).

We choose the eccentricity $e$ and inclination $i$ by the following
procedure.  From Fig.~3, we can estimate that a typical dust contour
has a ratio of distances closest and farthest from the core center
given in a model of nested confocal ellipses by $(1-e)/(1+e)\approx
0.30$, which implies $e \approx 0.54$.  Similarly, we may estimate that
these ellipses have an apparent minor-to-major axis-ratio of
$(1-e^2)^{1/2}\cos i \approx 0.54$, which implies $\cos i \approx
0.64$.  The resulting ellipses for three iso-surface-density contours,
spaced in a geometric progression 1:2:4, are shown as solid curves in
Fig.~3.

Determination of $\cos i$ allows us to compute an expected $\langle
B_\parallel\rangle$ = $\langle B_z \rangle\cos i$.  Similarly, we
obtain the expected hydrogen column density by multiplying $\langle
\Sigma \rangle$ by $(\cos i)^{-1}$ for a slant path through an inclined
sheet and by 0.7 for the mass fraction of H nuclei of mass $m_H$:
$N_H=0.7 \langle \Sigma \rangle / (m_H\cos i)$.

The sound speed for the 10 K gas in L1544 is $a=0.19$~km~s$^{-1}$
(Tafalla et al.~1998).  These authors give $\Delta V = 0.22$ km
s$^{-1}$ as the typical linewidth for their observations of C$^{34}$S
in this region.  For such a heavy molecule, turbulence is the main
contributor to the linewidth, which allows us to estimate the mean
square turbulent velocity along a typical direction (e.g., the line of
sight) as $v_t^2 = \Delta V^2/8\ln 2$.  We easily compute that $v_t^2$
has only 24\% the value of $a^2$.  Assuming that it is possible to
account for the ``pressure'' effects of such weak turbulence by adding
the associated velocities in quadrature, $a^2+v_t^2$, we adopt an
effective isothermal sound speed of $a=0.21$ km s$^{-1}$ for L1544.

The radius of the Arecibo telescope beam at the distance of L1544 is
$R=0.06$~pc (Crutcher \& Troland~2000).  Ambipolar diffusion
calculations by Nakano (1979), Lizano \& Shu (1989), Basu \&
Mouschovias (1994) suggest that $\lambda \approx 2$ when the pivotal
state is approached (see the summary of Li \& Shu 1996).  Putting
together the numbers, $\cos i = 0.64$, $R = 0.06$ pc, $a = $ 0.21 km
s$^{-1}$, and $\lambda = 2$, we get $\langle
B_\parallel\rangle=11$~$\mu G$, in excellent agreement with the Zeeman
measurement of Crutcher \& Troland (2000).  These authors also deduce
$N_H=1.8\times 10^{22}$ cm$^{-2}$ from their OH measurements, whereas
we compute a hydrogen column density within the Arecibo beam of
$N_H=1.4 \times 10^{22}$ cm$^{-2}$.  The slight level of disagreement
is probably within the uncertainties in the calibration or calculation
of the fractional abundance of OH in dark clouds (cf. Crutcher 1979,
van Dishoeck \& Black 1986, Heiles et al. 1993).

Our ability to obtain good fits of much of the observational data
concerning the prestellar core L1544 with a simple analytical model
should be contrasted with other, more elaborate, efforts.  Consider,
for example, the {\it axisymmetric} numerical simulation of Ciolek \&
Basu~(2000), who were forced to assume a disk close to being edge-on
($\cos i \approx 0.3$ when $e$ is assumed to be 0) to reproduce the
observed elongation, but who left unexplained the eccentric
displacement of the cloud core's center (very substantial for ellipses
of eccentricity $e \approx 0.54$).   The adoption of axisymmetric cores
leads to another problem:  Ciolek \& Basu's deprojected magnetic field
is on average 3-4 times stronger than ours, values never seen directly
in Zeeman measurements of low-mass cloud cores.  [See the comments of
Crutcher \& Troland (2000) concerning the need for magnetic fields in
Taurus to be all nearly in the plane of the sky if conventional models
are correct.]  Natural elongation plus projection effects, as
anticipated in the comments of Shu et al.~(1999), allow us to model
L1544 as a moderately supercritical cloud, with $\lambda \approx 2$,
fully consistent with the theoretical expectations from ambipolar
diffusion calculations, and in contrast with the value $\lambda \approx
8$ estimated by Crutcher \& Troland~(2000) from the 
measured values of $B_\parallel$ and $N_H$. In addition, if L1544 is a thin,
{\it intrinsically eccentric}, disk seen moderately face-on, as implied
by our model, then the extended inward motions observed by Tafalla et
al.~(1998; see also Williams et al.~1999) may be attributable to a
(relatively fast) core-amplification mechanism that gathers gas
(neutral and ionized) dynamically but subsonically along magnetic field
lines on both sides of the cloud toward the disk's midplane.

Finally, we show in Fig.~3 the direction of the average magnetic
field projected in the plane of the sky predicted by our model (thin solid
line) and derived from submillimeter polarization observations of
Ward-Thompson et al.~(2000) (thin dashed line). Since we have assumed
in our model that the major axis of iso-surface-density contours is in
the plane of the sky, the predicted projection of the magnetic field is
parallel to the cloud's minor axis. The offset between the measured
position angle of the magnetic field and the cloud's minor axis might
indicate some inclination of the cloud's major axis with respect to the
plane of the sky.  The turbulent component of the magnetic field, not
included in our model, may also contribute to the observed deviation.

\section{Conclusions}

We close with the following analogies.  The basic problem with trapped
magnetic fields is that they compress like relativistic gases (i.e.,
their stresses accumulate as the 4/3 power increase of the density in
3-D compression).  Such gases have critical masses [e.g., the
Chandrasekhar limit in the theory of white dwarfs, or the magnetic
critical mass of equation~(1)] which prevent their self-gravitating
collections from suffering indefinite compression, no matter how high
is the surface pressure, if the object masses lie below the critical
values.  Moreover, while marginally supercritical objects might
collapse to more compact objects (e.g., white dwarfs into neutron
stars, or cloud cores into stars), a single such object cannot be
expected to naturally fragment into multiple bodies (e.g., a single
white dwarf with mass slightly bigger than the Chandrasekhar limit into
a pair of neutron stars).

In order for fragmentation to occur, it might be necessary for the
fluid to decouple rapidly from its source of relativistic stress.  For
example, the universe as a whole always has many thermal Jeans masses.
Yet in conventional big-bang theory, this attribute did not do the
universe any good in the problem of making gravitationally bound
subunits, as long as the universe was tightly coupled to a relativistic
(photon) field.  Only after the matter field had decoupled from the
radiation field in the recombination era, did the many fluctuations
above the Jeans scale have a chance to produce gravitational
``fragments.'' It is our contention that this second analogy points
toward where one should search for a viable theory of the origin of
binary and multiple stars from the gravitational collapse of magnetized
molecular cloud cores.

\end{document}